 \documentclass[twocolumn,twoside,showpacs,preprintnumbers,superscriptaddress,amsmath,amssymb,prl]{revtex4}
 
 \usepackage{cases}
 \usepackage{CJK}
 \usepackage{txfonts}
 \usepackage{mathrsfs}
 \usepackage{amssymb}
 \usepackage[dvips]{graphicx}
 \usepackage{dcolumn}
 \usepackage{bm}
 \usepackage{subfigure}
 \usepackage{floatflt}
 \usepackage{float}
 \usepackage{rotating}
 \usepackage{multirow}
 \usepackage[bookmarksnumbered,bookmarksopen,colorlinks,citecolor=blue,linkcolor=blue]{hyperref} %
 \usepackage{booktabs}
 \usepackage{epsfig,graphics,psfrag,amsmath}
 \usepackage{sidecap}

\def\thu{Department of Physics, Tsinghua University, Beijing 100084, China}

\def\imp{Institute of Modern Physics, Chinese Academy of Sciences, Lanzhou 730000, China}
\def\hnu{Institute of Particle and Nuclear Physics, Henan Normal University, Xinxiang 453007, China}
\def\sinap{Shanghai Institute of Applied Physics, Chinese Academy of Science, Shanghai 201800, China}
\def\sari{Shanghai Advanced Research Institute, Chinese Academy of Science, Shanghai 201210, China}
\def\umcs{Institute of Physics, Maria Curie Sk{\l}odowska University, 20-031 Lublin, Poland}
\def\gxnu{College of Physics and Technology, Guangxi Normal University, Guilin 541004, China}
\def\gxkl{Guangxi Key Laboratory of Nuclear Physics and Technology, Guangxi Normal University, Guilin 541004, China}
\def\ciae{China Institute of Atomic Energy, Beijing 102413, China}
\def\huz{School of Science, Huzhou University, Huzhou, 313000, China;}
\def\cicq{Collaborative Innovation Center of Quantum Matter, Tsinghua University, Beijing 100084, China}
\def\sichuan{College of Physics, Sichuan University, Sichuan, 610065, China}

\def\esym{$E_{\rm sym}(\rho)$}
\def\amev{MeV/u}
\def\krpb{$^{86}$Kr+$^{\rm 208}$Pb}

\def \rthe{$R({\rm t/^3He})$}

\begin{document}
 \bibliographystyle{unsrt}
 \preprint{00-000}

\title {Observing the Ping-pong Modality of Isospin Degree of Freedom in Cluster Emission from Heavy Ion Reactions} 

 \author{Yijie Wang} \email{yj-wang15@tsinghua.org.cn}
 \affiliation{\thu}
 
 \author{Fenhai Guan}
 \affiliation{\thu}
 
 \author{Xinyue Diao}
 \affiliation{\thu}
 
 \author{Mengting Wan}
 \affiliation{\gxnu}
 \affiliation{\gxkl}
 
 \author{Yuhao Qin}
 \affiliation{\thu}
 
 \author{Zhi Qin}
 \affiliation{\thu}

\author{Qianghua Wu}
\affiliation{\thu}

\author{Dong Guo}
\affiliation{\thu}

\author{Dawei Si}    
\affiliation{\thu}

\author{Sheng Xiao}  
\affiliation{\thu}

\author{Boyuan Zhang}  
\affiliation{\thu}

\author{Yaopeng Zhang} 
\affiliation{\thu}

\author{Baiting Tian} 
\affiliation{\sichuan}

\author{Xianglun Wei}
\affiliation{ \imp }

\author{Herun Yang}
\affiliation{ \imp }

\author{Peng Ma}
\affiliation{ \imp }

\author{Rongjiang Hu}
\affiliation{ \imp }

\author{Limin Duan}
\affiliation{ \imp }

\author{Fangfang Duan}
\affiliation{ \imp }

\author{Qiang Hu}
\affiliation{ \imp }

\author{Junbing Ma}
\affiliation{ \imp }

\author{Shiwei Xu}
\affiliation{ \imp }

\author{Zhen Bai}
\affiliation{ \imp }

\author{Yanyun Yang}
\affiliation{ \imp }

\author{Jiansong Wang}
\affiliation{ \imp }
\affiliation{ \huz }

\author{Wenbo Liu}
\affiliation{\hnu }

\author{Wanqing Su}
\affiliation{\hnu }

\author{Xiaobao Wei}
\affiliation{\hnu }

\author{Chun-Wang Ma}
\affiliation{\hnu }

\author{Xinxiang Li}
\affiliation{\sinap}

\author{Hongwei Wang}
\affiliation{\sinap}
\affiliation{\sari}
 
 \author{Fangyuan Wang}
 \affiliation{\ciae }
 
 \author{Yingxun Zhang}
 \affiliation{\ciae }
 
 \author{Micha{\l} Warda}
 \affiliation{\umcs }

 \author{Arthur Dobrowolski}
 \affiliation{\umcs }

 \author {Bo{\.z}ena Nerlo-Pomorska}
 \affiliation{\umcs }
 
 \author{Krzysztof Pomorski}
 \affiliation{\umcs }
 
  \author{Li Ou} \email{liou@gxnu.edu.cn}
 \affiliation{\gxnu }
 \affiliation{\gxkl}

 \author{Zhigang Xiao } \email{xiaozg@tsinghua.edu.cn}
 \affiliation{\thu}
 \affiliation{\cicq}
 

 
 \date{\today}

 \begin{abstract}

  Two-body correlations of the isotope-resolved light and heavy clusters are measured  in \krpb~ reactions at 25 \amev. The yield and kinetic variables of the  $A=3$  isobars,  triton and $^3$He, are analyzed in coincidence with the heavy clusters of $7\le A \le 14$ emitted at the earlier chance.  While the velocity spectra of both triton and $^3$He exhibit scaling behavior over the type of the heavy clusters, the yield ratios of ${\rm t/^3He}$ correlate reversely to the neutron-to-proton ratio  $N/Z$ of the latter, showing the ping-pong modality of the $N/Z$ of emitted clusters. The commonality that the $N/Z$ of the residues keeps the initial system value is extended to the cluster emission in heavy ion reactions. The comparison of transport model calculations to the data is discussed.
    

 \end{abstract}


 \maketitle


  {\it Introduction -} 
  Since the detection of the gravitational wave from the neutron star merging event GW170817 \cite{LIGOScientific:2017vwq,LIGOScientific:2018cki,De:2018uhw}, it is more indispensable than ever to unveil the equation of the state (EOS) of the nuclear matter, which is essential to describe the explosion of the celestial objects where the heavy elements are created in the universe. Particularly, the EOS of asymmetrical nuclear matter, namely the density-dependent nuclear symmetry energy \esym, has been a longstanding  frontier of common interest bridging nuclear physics and astrophysics. Great progress has been made in the constraint of \esym~ using astrophysical observations and terrestrial experiments. However, there is still significant room to reduce the uncertainty \cite{Li:2021thg,Huth:2021bsp}.  
  
   In terrestrial nuclear laboratories,  depending on the beam energy, a variety of isobaric yield ratios in heavy ion collisions (HICs), like n/p, $\rm t/^3He $,  ${\rm \pi^-/\pi^+}$, $\rm K^0/K^+$ and $\rm \Xi^-/\Xi^0$ \cite{Li:1997rc,Zhang:2005sm,Li:2002qx,Xiao:2008vm,SRIT:2021gcy,Ferini:2006je,Yong:2022pyb,Chen:2003qj,Li:2005kqa},  have been identified to probe \esym~ in wide density range.  However, quantitative extraction of \esym~ is not trivial because of the complicated transport behavior of neutrons and protons in the collisions. To achieve precise results, it is essential to understand how the isospin degree of freedom (IDOF) evolves during the process of HIC \cite{Colonna:2020euy}. 
   
   Among the aforementioned isobaric ratios, the ratio of $A=3$ isobars, $\rm t/^3He$, has been suggested to probe the enriched feature of isospin dynamics in HICs both theoretically \cite{Chen:2003qj,Chen:2003ava,Wang:2014aba,Li:2005kqa,Gaitanos:2004zh,Yong:2009te,Chomaz:1998tp,Albergo:1985zz,Dempsey:1996zz} and experimentally \cite{FOPI:2010xrt,Xu:1999bs,Famiano:2006rb,Nagamiya:1981sd,Dempsey:1996zz,Veselsky:2000kp,FOPI:1999gxl,Gurov:2014hea}.  Nevertheless, more than inspecting the one-body quantity at final stage, it is recently realized that the temporal evolution of the isospin observable carries the fine effect of \esym~ \cite{Jedele:2017xow,RodriguezManso:2017emd}. Besides, isospin chronology reveals the particle emission hierarchy by measuring the correlation functions of like and unlike particle pairs \cite{Wang:2021mrv,Ghetti:2003zz, Ghetti:2003pv,Verde:2006dh}.
    
  Furthermore, the complication of clustering emerges. As reported recently, cluster formation on the surface of heavy nuclei depends on the $N/Z$ of the system and couples to the properties of dilute neutron-rich matter \cite{Tanaka:2021oll}.  As long as the cluster emission is abundant in HICs at intermediate energy, one may ask naturally, how does the IDOF evolve with the presence of cluster emission and is there any commonality of the relaxation of IDOF arising from the effect of \esym~ in a variety of nuclear process?
    
In this letter, we are motivated to differentiate the fine signal featuring the transport of IDOF related to the emission of clusters  and to identify a novel isospin probe using the temporal two-body correlation rather than a single isobaric yield ratio. The thermodynamic and chemical correlation of light and heavy clusters are analyzed in 25 MeV/u \krpb~ reactions.  The anti-correlation of the isobaric ratio $\rm t/^3He$ with  the  $N/Z$ of the heavier clusters are reported. Further, the commonality that the $N/Z$ of the initial system undergoing cluster emission is kept in various nuclear processes is discussed.
     
 {\it Experimental setup - }  The experiment was performed with the Compact  Spectrometer for Heavy IoN Experiment (CSHINE) \cite{Guan:2021tbi,Wang:2021jgu}, installed at the final focal plane  of  the Radioactive Ion Beam Line at Lanzhou (RIBLL-I) \cite{SUN2003496}. The $^{86}$Kr beam with 25 MeV/u incident energy was delivered by the Heavy Ion Research Facility at Lanzhou (HIRFL) \cite{XIA200211}, bombarding a natural lead target with the thickness of $1~ {\rm mg/cm^2}$. In the current configuration of CSHINE,  light charged particles (LCPs) and intermediate mass fragments (IMFs) are measured by 4 silicon-strip detector telescopes (SSDTs), while the coincident fission fragments (FFs) are measured by 3 parallel plate avalanche counters (PPACs), each covering a sensitive area of $240\times 280~ {\rm mm^2}$. Each SSDT consists of one single-sided silicon-strip detector (SSSSD) and one double-sided silicon strip detector (DSSSD), backed by a $3\times 3 $ CsI(Tl) crystal hodoscope with length of 50 mm. The sensitive area of each SSDT is ${\rm 64\times64 mm^2}$, and the granularity is $4\times 4~ {\rm mm^2}$ delivering about $1^\circ$ angular resolution. The SSDTs cover the angular range from $10^\circ -60^\circ$ in the laboratory system. The energy resolution of the SSDT is better than $2\%$, and the isotopes up to $Z=7$ can be identified. For the details of the construction of the CSHINE, one can refer to \cite{Guan:2021tbi,Wang:2021jgu}. Since the event statistics of the 4-body coincidence, i.e., with 2 charged particles in SSDTs and 2 FFs in PPACs, does not suffice, here we  analyze the two-body coincidence of two fragments in the SSDTs. Multi hits and signal sharing are treated with care in the track recognition, the overall track recognition efficiency is about 90\%. The detailed procedure of track reconstruction can be found in \cite{Guan:2021nfk}.

\begin{figure}[h] 
 \centering
 \includegraphics[angle=0,width=0.45\textwidth]{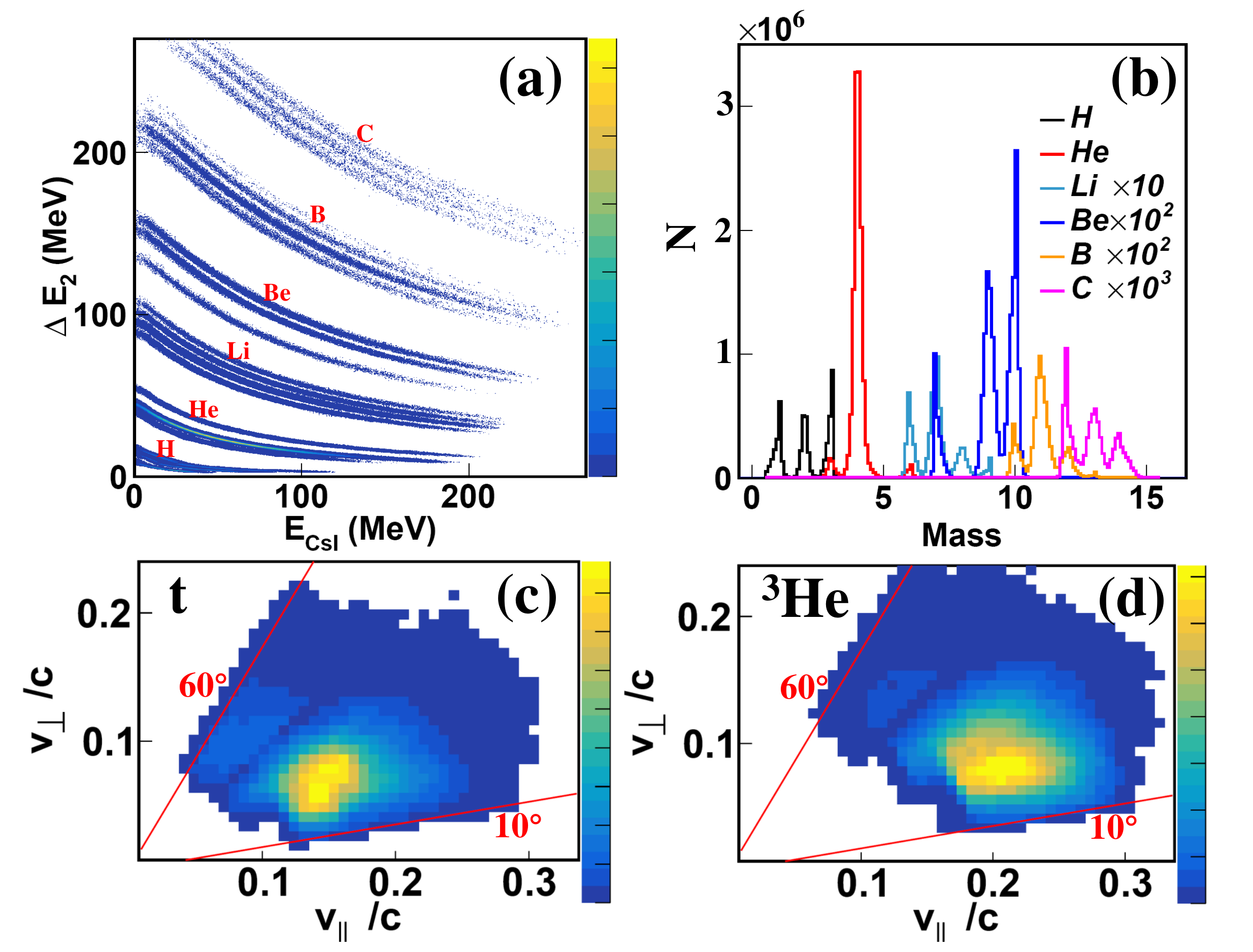}
 \caption{(Color online)  The scattering plot of $\Delta E_2 - E_{\rm CsI}$ of the SSDT1 (a). The PID plots for all isotopes on  $\Delta E_2 - E_{\rm CsI}$  summed over SSDT 1 to 4 (b). The phase space distribution of triton (c) and $^3$He (d) deposited in CsI(Tl) units.}
 \label{phase-space}
 \end{figure}

Fig. \ref{phase-space} summarizes the data used in the current analysis. Panel (a) presents the scattering plot of $\Delta E_2 - E_{\rm CsI}$ of the SSDT1. It is shown that all the isotopes can be identified clearly to $Z=6$. Panel (b) presents the PID plots for all isotopes on $\Delta E_2-E_{\rm CsI}$ summed over SSDT 1 to 4 after the tracking procedure \cite{Guan:2021nfk}. Panels (c) and (d) present the phase space distribution of triton and $^3$He stopped in the CsI(Tl) hodoscope, respectively.

{\it Results and Discussions -} 
In order to observe the temporally correlated emissions of charged particles, we investigate the two-body events in the SSDTs, with one being the light charged particle, ${\rm F_L}(Z_{\rm L},A_{\rm L})$, the other being a heavy fragment, ${\rm F_ H}(Z_{\rm H}, A_{\rm H})$, with $7 \le A_{\rm H} \le 14$, where the subscripts L and H denote light and heavy respectively. Here ${\rm F_ L}$ is chosen as one of the mirror nuclei, triton or $^3{\rm He}$. According to earlier studies using correlation function method, the heavy fragments are emitted with an averagely shorter time constant at this energy region \cite{YongHe:1998zz}. Thus, it is reasonable to assume that the ${\rm F_H}$ is emitted earlier than ${\rm F_L}$ in our study.  To eliminate the influence from the low energy statistical decay, we count only the high energy particles, i.e., only the ${\rm F_L}$ stopped in the CsI(Tl) hodoscope are analyzed. 

\begin{figure}[h] 
 \centering
 \includegraphics[angle=0,width=0.45\textwidth]{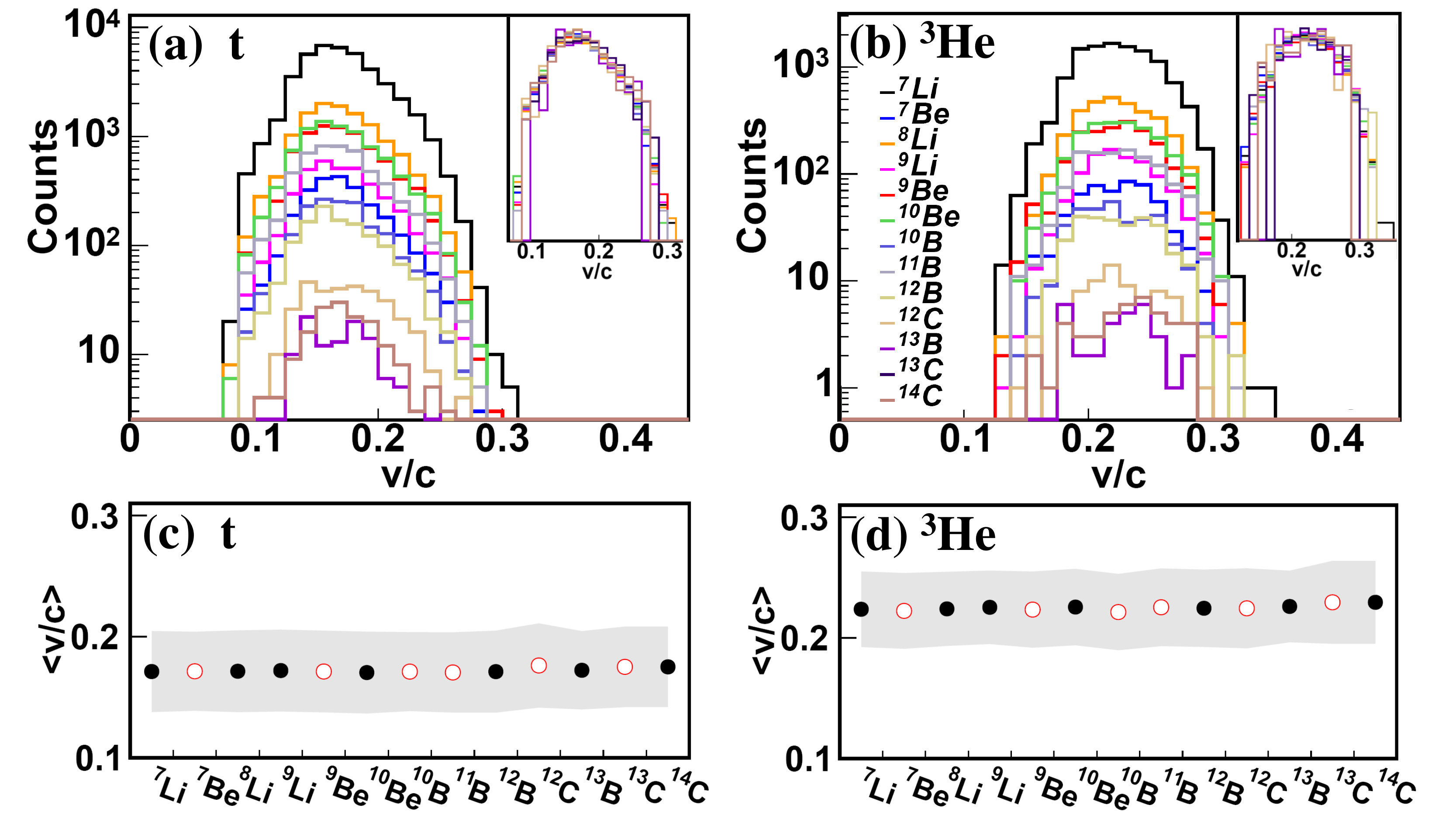}
 \caption{(Color online)  The spectra of the velocities (in unit of $c$) of triton (a) and $^3$He (b)  in coincidence with various $F_{\rm H}$ of $A_{\rm H}=7$ to 14. The lower panels present the mean velocity $\left<v\right>$ of t (c) and  $^3$He (d), with the solid (open) circles denoting the coincident $F_{\rm H}$ of larger (smaller) neutron richness,  respectively. The shadowing band denotes the standard deviation of each corresponding velocity spectrum. }
 \label{kinetic}
 \end{figure}

We first check the thermodynamic correlation. Fig. \ref{kinetic} (a) and (b) present the velocity spectra of triton (a) and  $^3$He (b)  in coincidence with various heavy fragments ${\rm F_H}$. It is shown that the shape and the range of the velocity distributions of triton and $^3$He exhibit insignificant dependence on the type of ${\rm F_H}$, although the coincident yield differs. Normalizing the yields to the black histograms corresponding to the coincidence with ${\rm ^7Li}$, one sees that all the spectra exhibit scaling behavior, as shown in the insets in (a) and (b). Further, the  mean velocity  $\left<v\right>$ of triton and  $^3$He  are presented in (c) and (d), varying with the type of $\rm F_H$. The shadowing band represents the standard deviation $\sigma_v$ of each individual velocity distribution. The solid and open symbols denote the ${\rm F_H}$ with larger and smaller neutron richness, respectively.  Clearly, both $\left<v\right>$ and $\sigma_v$ are nearly constant, showing insignificant dependence on the $N/Z$ of the heavy fragments.  It suggests that the light particles experience the same dynamic process, regardless of whether the heavy clusters are rich or deficient in neutrons.
  
\begin{figure}[h]
 \centering
 \includegraphics[angle=0,width=0.45\textwidth]{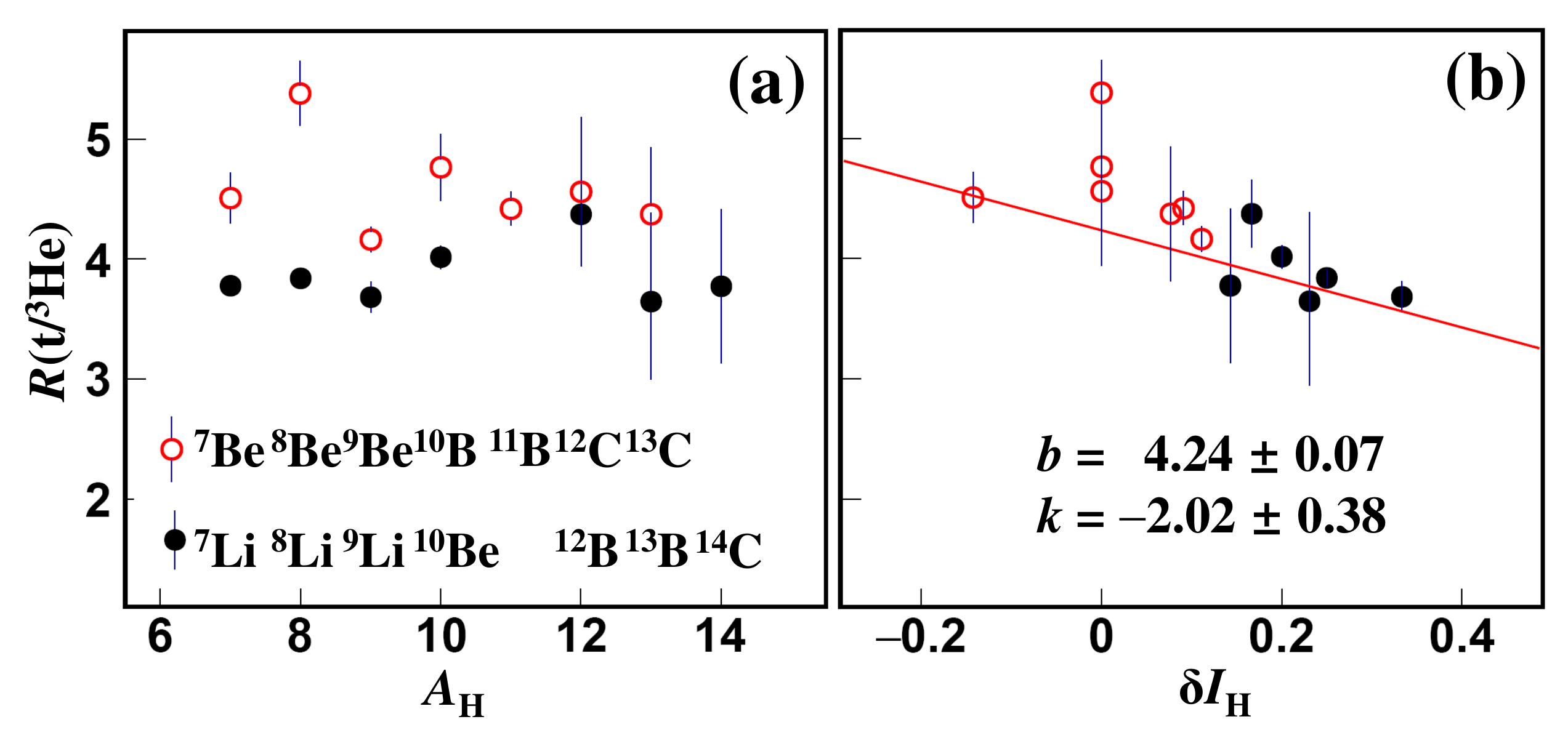}
 \caption{(Color online)  (a) Yield ratio of \rthe~ in coincidence with different heavy fragments, the solid (open) circles represent the heavy fragments of larger  (smaller) neutron richness at a given mass number.  (b) Yield ratio of \rthe~  as a function of the neutron richness $\delta I_{\rm H}$ of the heavy fragment $F_{\rm H}$.  $k$ and $b$ are the slope and the intercept parameters of the linear fit, respectively. }
 \label{ratio}
 \end{figure}

Surprisingly, however, the IDOF features differently. Fig. \ref{ratio} (a) presents the yield ratio of $\rm t/^3He$, written as \rthe,  as a function of the mass of  $F_{\rm H}$. Again, the solid (open) symbols correspond to the more (less) neutron-rich  ${\rm F_H}$ at a given mass. Here the heavy cluster ${^8}$Be is reconstructed by two correlated $\alpha$ particles. It is seen that the ratio \rthe~  splits in two groups, with an exception of the  ${\rm F_H}$ pair with $A=12$ where the effect is blurred by  large uncertainty. A ping-pong 
motion modality of the $N/Z$ in cluster emission is suggested: after a neutron-rich heavy fragment (solid) is emitted, the \rthe~ is smaller, indicating that a less neutron-rich ${\rm F_L}$ follows. Oppositely, if the ${\rm F_H}$ is less neutron rich (open), \rthe~ is larger, indicating that ${\rm F_L}$ is averagely more neutron-rich. The ping-pong behavior of the $N/Z$ of the two clusters demonstrates that the isospin content of  ${\rm F_L}$ depends on the $N/Z$ of ${\rm F_H}$ emitted at earlier chance.  This tendency is further presented in panel (b), where the \rthe~ is plotted as a function of the relative neutron richness of ${\rm F_H}$  defined by $\delta I_{\rm H} = (N_{\rm H}-Z_{\rm H})/A_{\rm H}$. The solid  line is the least square linear fit to the data points. A non-zero minus slope $k$ is convinced at $5.3\sigma$ confidential level, evidencing that the isospin compositions of  ${\rm F_H}$  and  ${\rm F_L}$ are anti-correlated. 

Model independently, the anti-correlation between the $N/Z$ of ${\rm F_L}$ and ${\rm F_H}$ infers  how the IDOF evolves towards equilibrium during the decay of the highly excited system formed in \krpb~ reactions. If more (less) neutrons are carried away by the heavy cluster at an earlier stage, the following light one carries less (more) neutrons. It suggests that the neutron richness of the emitted clusters is balanced near a certain value, so that the remaining system, possibly destined to emission residue or fission, possesses a certain $N/Z$ close to that of the initial system, see below.        

\begin{figure}[h]
 \centering
 \includegraphics[angle=0,width=0.45\textwidth]{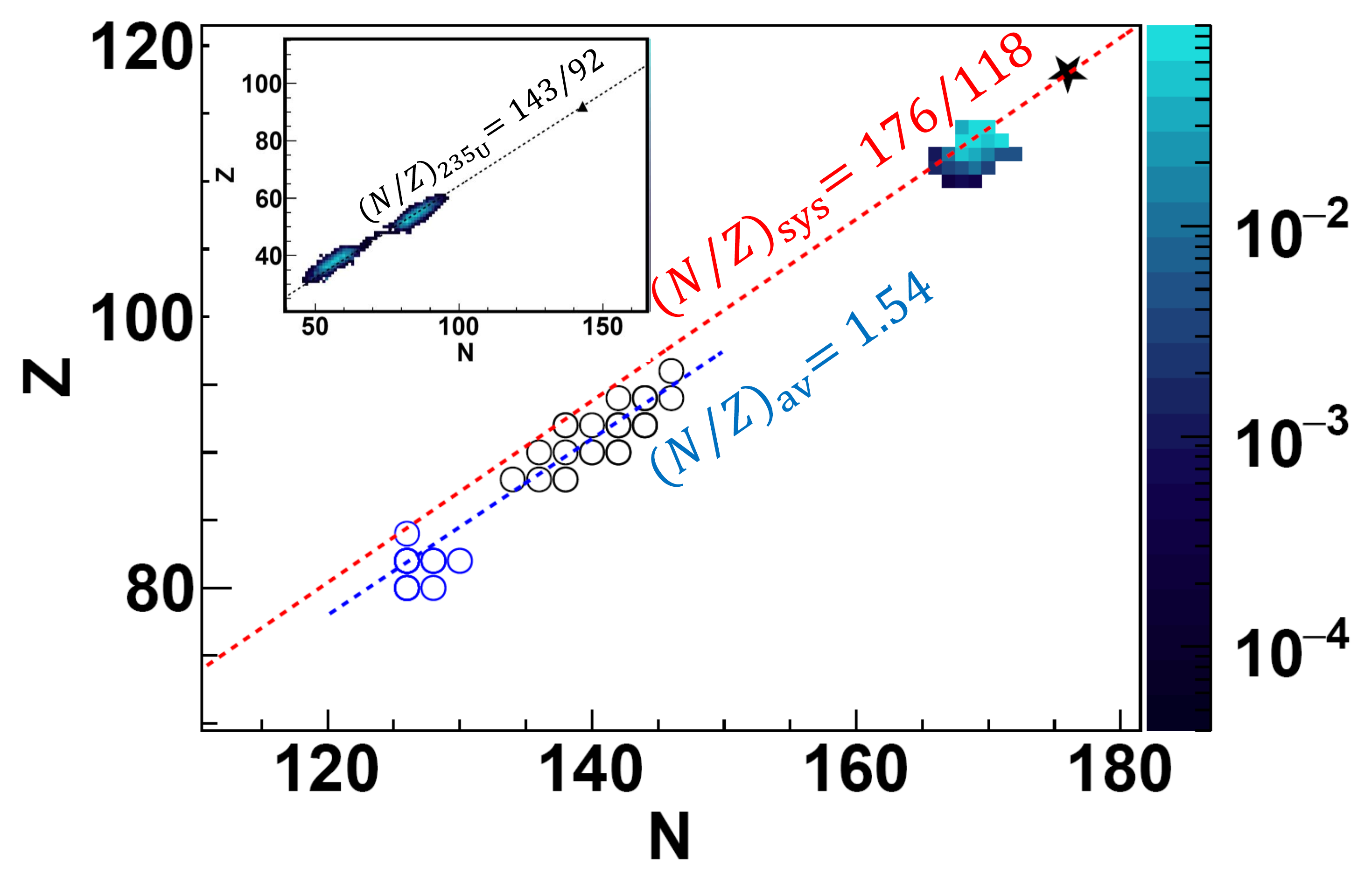}
 \caption{(Color online)  The correlation between the  $N/Z$ of the initial system and the remaining nuclei. The asterisk denotes the initial \krpb~ system, and the colored contour represents the probability distribution of the remaining nuclei after the emission of ${\rm F_ H}$ and ${\rm F_L}$. The circles denote the cluster decay of various super-heavy nuclei, with the black (blue) ones being the parents (daughters). The inset displays the yield distribution of the n-induced fission of $^{235}$U on the $N-Z$ plot.}
 \label{nz2frag}
 \end{figure}

In order to show the effect clearly, Fig. \ref{nz2frag} presents the $N-Z$ correlation of the remaining nuclei, obtained by subtracting ${\rm F_L}$ and ${\rm F_H}$ from the initial system. The $N/Z=1.497$ relationship is indicated by the red dashed line, and the asterisk denotes the location of  \krpb~ system.  It is shown that the remaining system situates in the vicinity of the dashed line, indicating that the average $N/Z$ of the remaining system is the same as the initial value of \krpb.  Here we can only count the two clusters measured, including $A=3$ isobars and the heavy clusters. Because of the anti-correlation demonstrated in Fig. \ref{ratio}, it is reasonable to assume the undetected decay particles have a similar $N/Z$ to the detected products and thus do not drag the distribution center away from the dashed line.

Interestingly, the cluster radioactivity observed in various super-heavy nuclei \cite{Rose:1984zz,Poenaru:2002zp,Warda:2018zdm} is similar, as shown by the circles in the plot. The cluster decay channels are taken from \cite{Poenaru:2002zp}. It is seen that  the daughters (blue) are distributed near the  same $N/Z$ line of the parent nuclei (black). Moreover, if one inspects the fission of heavy nuclei, for instance, the neutron induced fission of $^{235}$U plotted in the inset, the $N/Z$ of the fission fragments follows the same $N/Z$ value of the parent nucleus $^{235}$U \cite{NNDC}. Because of the presence of \esym, which  finally goes into the Q value, it is seemingly common that the $N/Z$ of the daughters inherits that of the parents in fission and in the cluster decay of heavy nuclei. Our experiment extends the commonality to the cluster emission of the highly excited system formed in heavy ion reactions.

To understand  the isospin transport behavior, the reaction process is simulated using the improved isospin-dependent quantum molecular dynamics model (ImQMD05) \cite{Zhang:2020dvn} followed by a statistic decay afterburner GEMINI \cite{Charity:1988zz, Charity:2001yu}. 
The local nuclear potential energy density functional in the ImQMD05 model is written as 
    \begin{align}\label{Vloc}
	V_{\rm{loc}}&=\frac{\alpha}{2}\frac{\rho ^{2}}{\rho _{0}}+\frac{\beta }{\eta +1}%
	\frac{\rho ^{\eta +1}}{\rho _{0}^{\eta }}+\frac{g_{\rm{sur}}}{2\rho _{0}}\left(\nabla \rho \right)^{2}\\ \nonumber
	&+\frac{g_{\rm{sur,iso}}}{\rho_{0}}[\nabla(\rho_{\rm n}-\rho_{\rm p})]^{2} 
	+g_{\rho\tau}\frac{\rho^{8/3}}{\rho_{0}^{5/3}}+\frac{C_{\rm s}}{2} \frac{\rho^{\gamma+1}}{\rho_0^{\gamma}} \delta^{2},
\end{align}
where $\rho$, $\rho_{\rm n}$, $\rho_{\rm p}$ are the nucleon, neutron, and proton density, $\delta = (\rho_{\rm n}-\rho_{\rm p})/(\rho_{\rm n}+\rho_{\rm p})$ is the isospin asymmetry. $\rho_0$ is the saturation density. The parameters in Eq. \eqref{Vloc} are determined by Skyrme interaction with parameter set MSL0 \cite{Chen:2010qx} except for $C_{\rm{s}}=36.0~\rm MeV$ and the variable $\gamma$ relevant to \esym. The density behavior of \esym~ can be mimicked by varying $\gamma$.

\begin{figure}[h]  
 \centering
 \includegraphics[angle=0,width=0.45\textwidth]{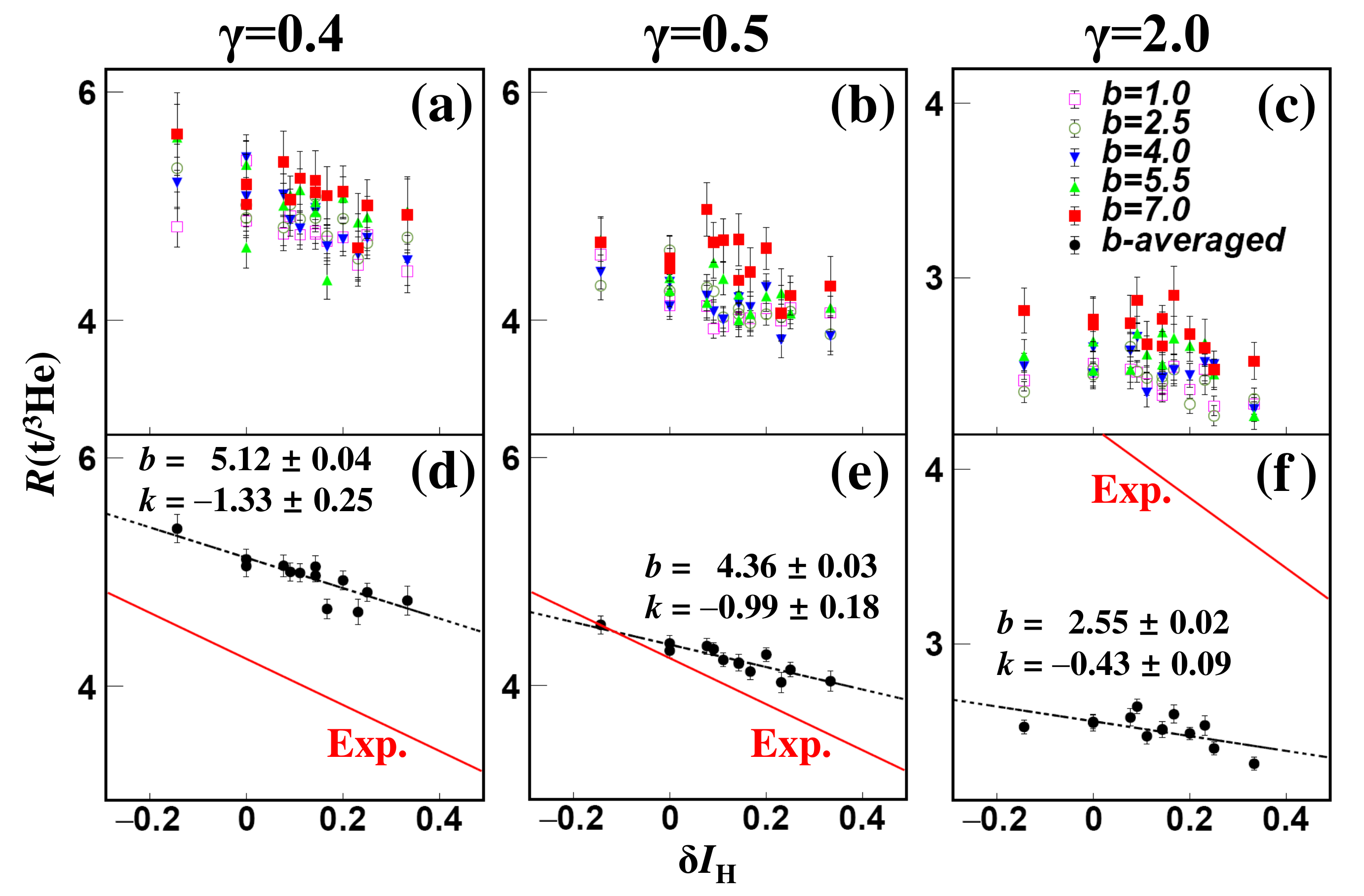}
 \caption{(Color online) Yield ratio of \rthe~  by ImQMD simulations as a function of $\delta I_{\rm H}$ with $\gamma=0.4$, 0.5, and 2, respectively. The results with  $b=1$ to 7 fm are presented in panels (a-c). The $b-$averaged results are in panels (d-f) in comparison to the experimental trend.}
 \label{imqmd1}
 \end{figure}

The calculations cover the impact parameter range of $1\le b \le 7$ fm. The same cuts on phase space are adopted as in experiment. To avoid the ambiguity of the correlation, we exclude the events containing more than one ${\rm F_H}$ with the same $N_{\rm H}$ and $A_{\rm H}$, but set no cut on the multiplicity of $F_{\rm L}$.  Fig. \ref{imqmd1} presents the \rthe~ ratio as a function of $\delta I_{\rm H}$ for different \esym~ with $\gamma=0.4$ (super-soft), 0.5 (soft), 2.0 (super-stiff), respectively. The simulated ratios depend slightly on  $b$, but significantly on the stiffness of \esym, as shown in panels (a-c). Panels (d-f) present the $b$-averaged results. The dashed lines are the linear fit, and the experimental trend is replotted in solid lines for comparison.  Interestingly, the anti-correlation between \rthe~ and $\delta I_{\rm H}$ is qualitatively reproduced with a soft \esym~($\gamma=0.5$).  However, the super-stiff and the super-soft  \esym~ are disfavored by the significant deviation from the experimental trend. Since the discrepancies, particularly on the slope, still exist between the data and the simulations, we do not quantify \esym~ here. The modeling of clustering in transport theory requires further efforts, even for the $A=3$ clusters, although \rthe~ has been  proposed to probe \esym~ in heavy ion reactions.

 {\it Summary -}   The $A=3$ isobars are measured in coincidence with a heavy cluster of $7\le A \le 14$  in 25 MeV/u \krpb~ reactions.  While the velocity spectra of triton and $^3$He show scaling behavior over the type of heavy cluster, the yield ratio \rthe~ exhibits evident anti-correlation with the $N/Z$ of the heavy clusters, suggesting the ping-pong modality of the $N/Z$ of the emitted clusters. The commonality  that the remaining system inherits the $N/Z$ of the initial system, which has been reported in cluster decay and fission of super heavy nuclei, is extended to the cluster emission from HIC at Fermi energies. In addition to the yield ratio of ${\rm  t/^3 He}$, the anti-correlation between the $N/Z$ of the two clusters provides a new line to test the transport theory in terms of cluster formation and isospin dynamics, and to  eventually achieve the accurate constraint of \esym~ using heavy ion reactions.   
 

{\it Acknowledgement -}  This work is supported by the National Natural Science Foundation of China under Grant Nos.
 11875174, 11961131010, 11961141004 and 
  11965004, 
  by the Ministry of Science and Technology under Nos. 2020YFE0202001 and 2022YFE0103400, 
  by the Polish National Science Center under No. 2018/30/Q/ST2/00185, 
  and by Tsinghua University Initiative Scientific Research Program and the Heavy Ion Research Facility at Lanzhou (HIRFL). 
  The authors thank Prof. Pengfei Zhuang from THU, Yueheng Lan from BUPT and Prof. Nobuaki Imai from Univ. Tokyo for their valuable discussions.

 \bibliography{refs}


\end{document}